# THE FORMATION OF WELL-DEFINED CRYSTALLINE STRUCTURES BY UV LASER IRRADIATION OF AMORPHOUS SILICON FILMS

Saydulla K. Persheyev[1,2], James A. Cairns[2]

Nanophotonics Laboratory, School of Physics and Astronomy, St Andrews University,
St Andrews, UK
Carnegie Physics Laboratory, Electronic Engineering Physics and Renewable Energy,
University of Dundee, Dundee, UK

sp95@st-andrews.ac.uk, tel: +44-1334461653

**Abstract**

This study provides a new insight into the processes which occur when thin film hydrogenated amorphous silicon is subjected to UV laser irradiation in the presence of oxygen. It achieves this by observing the effects of subjecting the films to progressively increasing laser radiation doses. This reveals that an array of nuclei is first created, leading to the formation of a well-defined crystalline network, consistent with the structure of silicon oxide. Further irradiation results in the formation of cone-like structures on the crystalline network, due to silicon-oxygen bond breakage and migration of the resultant silicon-rich material. Eventually the cone-like structures become the dominant features, but remain interconnected by nanowire remnants of the original crystalline structure.

Excimer laser processing of amorphous hydrogenated silicon (a-Si:H) has a long history of attracting scientific attention because if its potential for various applications, such as thin film transistors (TFT) [1], solar cells [2,3], and field emission devices [4]. Among the main attractions of a-Si:H is its mature technology; its use of inexpensive materials; large area capability; low temperature deposition (typically 200°C); and variable substrate choices, including plastics and flexible materials. Excimer laser technology using 248nm wavelength and 20ns pulse width, with a homogenised large area beam is also one of the conventional and industrially used methods for thin film processing. Excimer laser ultraviolet radiation is highly absorbed in amorphous hydrogenated silicon due to a specific optical gap (1.8eV), which allows it to modify amorphous films to produce complex nanocrystalline silicon grain-embedded structures with conical needle-like features [5], capable of absorbing light over a wide range of optical spectra [6].

Despite silicon nanowires (SiNWs) becoming one of the most important components of future electronic and sensor devices, there remain significant barriers to their implementation, such as the requirement for their deposition to be conducted at high temperature, and the need to use metal catalysts. The metal catalyst most commonly used is gold and it continues to dominate, [7,8] in spite of gold's well-known propensity to create traps for electrons and holes, thereby leading to uncontrollable device doping. An alternative catalyst is aluminium, which has been used to achieve lower temperature silicon nanowire deposition via a vapour–solid–solid mechanism [9]. In an alternative approach, room temperature growth of SiNWs (40-200nm) has been achieved by electroless etching of a silicon wafer in an HF solution containing Ag atoms (Ag-induced selective etching) [10]. The versatility of SiNWs has been extended by the discovery that they can be modified with different antibody receptors to selectively recognise different biological and chemical species. Consequently, Nanowire Field Effect Transistors (NW-

FETs) can be configured as ultrasensitive, selective sensors. The nanowire conductance is altered by the presence of molecules, for example proteins, bound to the receptor on the nanowire surface [11-13].

Excimer laser processing of amorphous hydrogenated films with thicknesses of 100-500nm has been studied extensively [14-16]. Vertically aligned amorphous hydrogenated silicon NWs and nanocones (NCs) have been produced by Reactive Ion Etching, using $SiO_2$ nanoparticles as a mask.[17]. It has been shown [18] that silicon oxide may play an important role in Si nanowire growth. In this work, the synthesis of Si nanowires by laser ablation and thermal evaporation was achieved by using pure Si powder mixed with $SiO_2$. The presence of the Si oxide was a key factor in enhancing the nucleation and one-dimensional growth of the Si nanowires.

The aim of this study was to investigate excimer laser processing of very thin (<50nm) amorphous hydrogenated silicon films, including the influence of laser processing parameters on the film topography, and microstructure.

Intrinsic non-doped a-Si:H films were deposited by Plasma Enhanced Chemical Vapour Deposition (PECVD) [5] on to low sodium Corning glass substrates. The silicon precursor gas was silane ($SiH_4$), introduced into the deposition chamber. The film growth time did not exceed 10 minutes. Before film deposition, the vacuum chamber was evacuated to $10^{-5}$ Torr. An argon plasma was used to clean both the substrate surface and the chamber, prior to deposition.

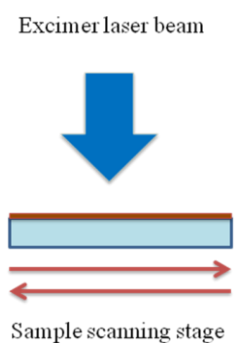

The amorphous silicon film irradiation was performed using a multiple-pulse 20ns KrF (248nm) excimer laser, with a spot size of 8mm by 4mm, and laser beam fluences of 140-300mJ/cm$^2$, which is near to the silicon ablation threshold. The laser beam repetition rate was 50Hz. The a-Si:H samples were scanned during laser annealing at 1mm/s , as described in detail elsewhere [14].

Fig.1 Excimer laser processing of amorphous silicon films on glass substrates

Scanning Electron Microscopy (SEM) of the surface topography was carried out using a Philips XL30 ESEM with a field emission electron gun. Samples were coated with a 2nm gold layer prior to imaging. The electron acceleration voltage was 15kV.

The amorphous hydrogenated silicon films were found to have a highly uniform topography, with a surface roughness not exceeding 1.8nm, as confirmed by AFM. The laser energy applied to the film is expected to be highly absorbed at the surface, causing crystallisation to occur at the appropriate temperatures. It was anticipated that this crystallisation process would be followed inevitably by silicon surface modification. It was observed in fact that laser fluences below 122mJ/cm$^2$ did not have any morphological effect on the films, but at fluences of 122mJ/cm$^2$ and above a dramatic change in film morphology and material transformation occurred. The initial nucleation (fig.1a) progressed as the laser fluence was increased. At 140mJ/cm$^2$ (fig. 1b) increasing grain size was observed, and at 156mJ/cm$^2$ (fig. 1c) silicon-free areas became apparent, due to the formation of net-like structures. On increasing the fluence to 177mJ/cm$^2$ (fig.1d) it was clear that the net-like structures were heavily decorated with cone-like growths. When the fluence was increased to 213mJ/cm$^2$ (fig.1e) these cone-like growths became even more predominant. This process is shown in more detail in Fig. 3(a) and (b), where the cone-like structures are clearly interlinked by silicon nanowires. A further increase in the fluence to 233mJ/cm$^2$ resulted in the silicon cone-like structures becoming deformed into more irregular shapes and the Si nanowires becoming discontinuous and damaged.(fig.2f).

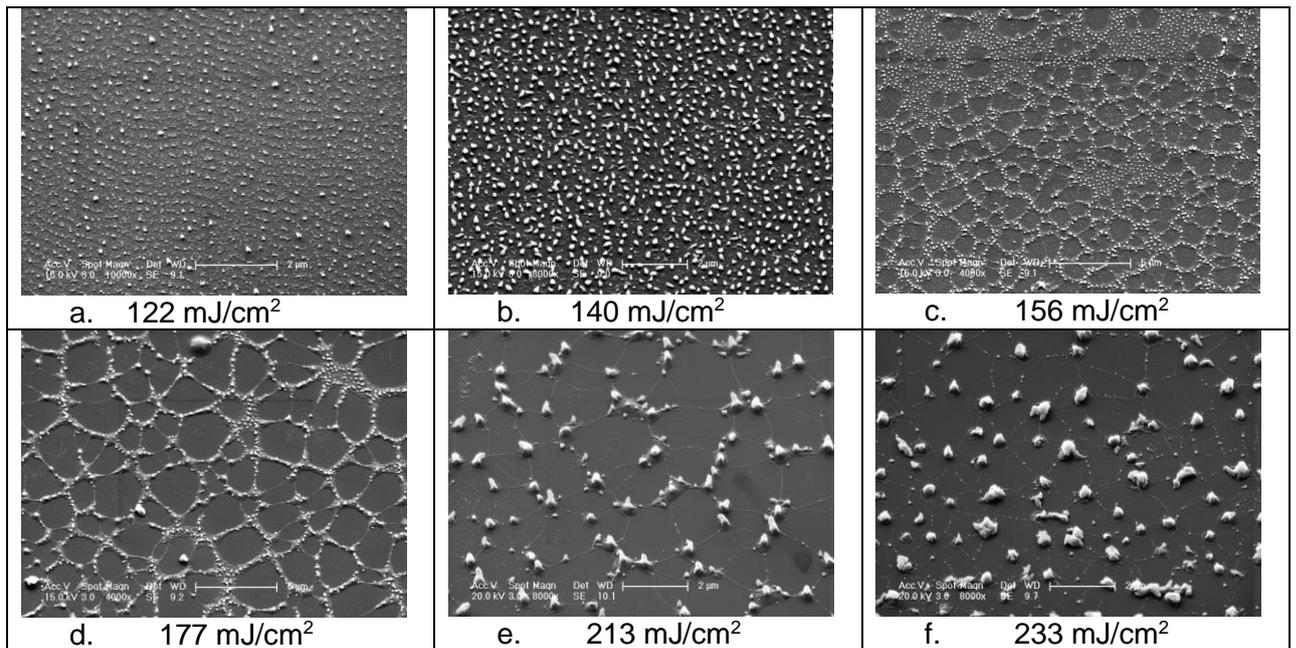

Figure 2. SEM images of laser processed thin silicon films at different beam fluences.

Higher magnification images of the silicon structures processed at laser fluences of 177 and 213mJ/cm$^2$ are shown in figures 3(a) and (b) respectively. In fig.3(a) cone-like structures (around 500nm in height) are seen, distributed within a network. On increasing the laser energy further, the cones are observed to have become more separated, *but linked by filament-like silicon nanowires* (b).

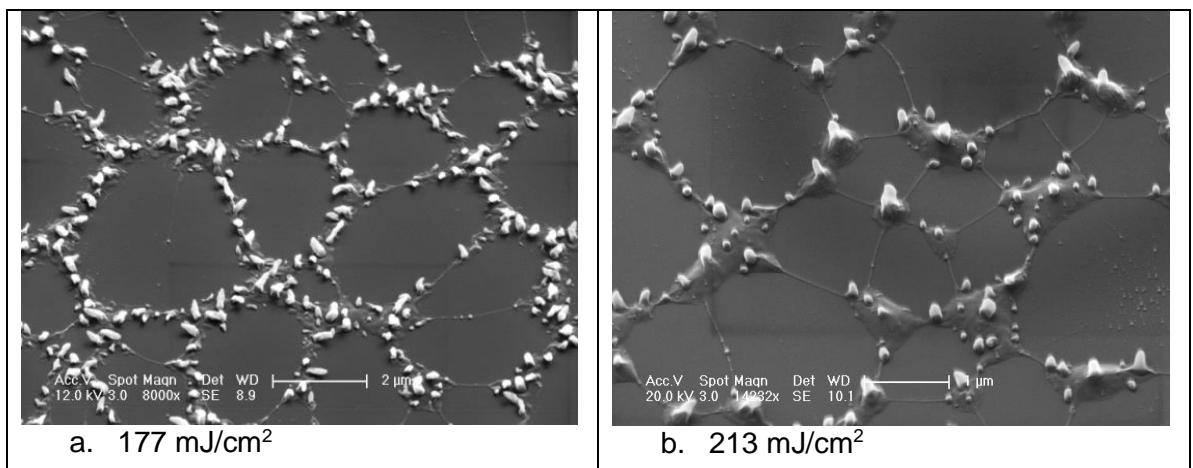

Figure 3. SEM images of laser processed thin silicon films at two different laser beam fluences.

Further SEM imaging reveals that the single conical structures are associated with three or four nanowire bridging connections (fig.4a). It is interesting to note that the nanowire to nanowire connections occur in a well-defined manner, such as the 'Y' shape shown in fig.4 b.

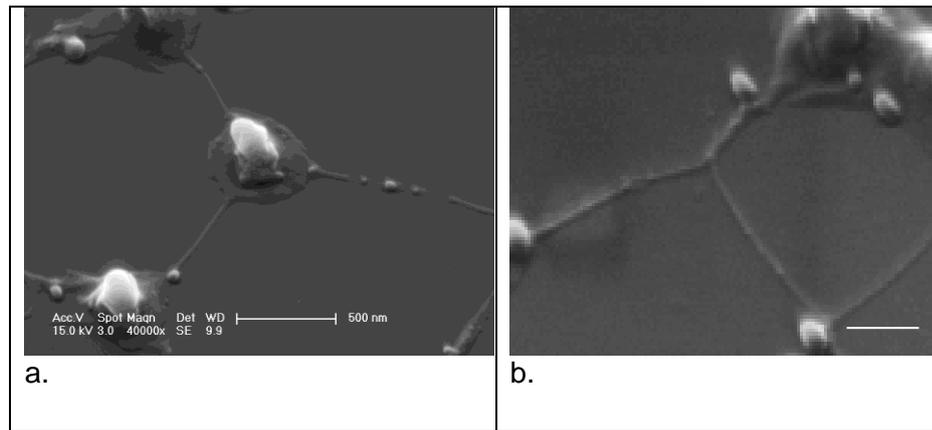

Figure 4. SEM (tilted 35 degrees) images of conical spike-like structures, linked to each other by a network of silicon nanowires; a: single cone-like structure with three silicon nanowire connection, b: 'Y' -type nanowire assembly (200 nm scale bar).

It is suggested that the above observations can be rationalised in the following way. First, UV-irradiation of the amorphous silicon film results in the creation of nodular features (Fig 2a and b). Subsequent higher dose irradiation results in the creation of a well-defined crystalline structure, shown in Figs2c and d. This structure is consistent with the tetragonal crystalline form expected of crystalline silicon oxide. It should be noted that in this study it forms only in the presence of ambient oxygen. No such effect was seen when the irradiation was conducted in vacuum ($10^{-3}$ Torr), or in the presence of nitrogen or sulphur hexafluoride. Closer inspection of the crystalline structure shows that it is decorated with nodules (Fig 3a and b). A possible explanation for this is that the UV irradiation has caused the silicon oxide to decompose, thereby creating silicon-rich nodules. This decomposition mechanism is discussed by Wang et al [18]. Subsequently, on subjecting the film to even higher dose irradiation, these cone-like nodules become more pronounced, possibly as they migrate over the surface and form larger clusters by a process such as Ostwald ripening [19]. It is interesting to note that evidence of the original crystalline network remains in place, as may be seen in Figs 4a and b, presumably as a consequence of the strong bonding at the interface between the silicon oxide and the glass substrate.

**Acknowledgments**
The authors wish to thank the following Dundee University colleagues: Mr Stuart Anthony, for growing the amorphous silicon films; Dr Y.Fan for the excimer laser processing; and Mr Martin Kierans for help in obtaining the high resolution SEM images.